\documentclass[12pt]{article} 
\title{Spin 1/2 and Invariant Coefficients}
\author{{\it Richard Shurtleff~}\thanks{affiliation and mailing 
address: Department of Sciences, 
Wentworth Institute of Technology, 550 Huntington Avenue, 
Boston, MA, USA, 02115, telephone number: (617) 989-4338, fax 
number: (617) 989-4591 , e-mail address: shurtleffr@wit.edu}} 
\begin{document} 

\maketitle 
\begin{abstract}
Massive spin 1/2 particles require 2-spinors for rotations, 4-spinors for rotations and boosts with parity. Including translations requires 8-spinors. Adapting 4-spinor field theory to 8-spinor fields with translation symmetry is discussed here. It is shown that four of these spin components act like conventional 4-spinors, satisfying the conventional free-particle Dirac equation. The remaining four components have unconventional coordinate dependence due to translation symmetry. One finds that the 4-vector current of the 8-spinor is proportional to the electromagnetic 4-vector potential of the conventional 4-spinor, with arbitrary charge. Thus translation symmetry in field theory unites a conventional 4-spinor field and its electromagnetic vector potential.


\end{abstract}

\pagebreak
\section{Introduction}

Relativity theory tells us that the essential aspects of a physical explanation of an experiment's results can be expressed in any inertial coordinate system however it is rotated, boosted or translated. Nevertheless, it is conventional to consider only quantum fields that are scalars under translations,\cite{Ramond p 19 (4.5), implicitly W chapter 5} not allowing multicomponent fields to mix components under translations while promoting spin, the mixing of components when the system is rotated or boosted. Dismissing translations by considering only quantities such as quantum fields that are scalar under translations robs field theory of a source of relationships which may be potentially useful in explaining physical phenomena. Some rudimentary considerations are presented in this paper.

Translations differ from rotations and boosts by being inhomogeneous. Rotating or boosting a null 4-vector regurgitates the same null 4-vector, but a translation adds the same displacement to the null coordinates of the origin that it adds to the coordinates of any other point. This is just one example of the general donor/receiver situation for general spins. The coordinates receive a displacement from some donor which, in this case, is a scalar. 

For general spins, translations change spin components of a `receiver' field that gets a donation from a `donor' field. The spins of receiver and donor are said to be `linked'.\cite{Lyubarskii p310 after (74.4)} In current terminology, spins $(A,B)$ and $(C,D)$ can be linked when $A$ and $C$ differ by $\pm 1/2$ and also $B$ and $D$ differ by $\pm 1/2.$ Thus, in the above example, coordinates are the receiver 4-vector field, spin $(1/2,1/2)$, and the displacement is donated by some scalar, a spin $(0,0)$ field, with $(1/2,1/2)$ linked to $(0,0).$

Four-component quantum fields, 4-spinors, describe the rotation and boost behavior of massive spin 1/2 particles invariant under parity. With translations the donor/receivers can be set so that one 4-spinor donates to a recipient 4-spinor. An 8-spinor field is required with parity. It is shown here how the donor 4-spinor obeys the Dirac equation while the recipient 4-spinor receives contributions that complicate its part of the Dirac equation. However, the contributions can be interpreted along a different aspect of spin 1/2 particles. The contributions to the recipient 4-spinor produce a current that is shown to be proportional to the electromagnetic vector potential with the donor 4-spinor as a source. Aside from a scale factor, the electromagnetic vector potential of the 4-spinor donor field is the current of the 8-spinor donor/receiver field.

Sec. 2 introduces an 8-spinor representation of the Poincar\'{e} algebra of rotations, boosts and translations. In Sec. 3, the spin 1/2 field is deduced following a standard process\cite{Weinberg QTF I Chap5} from the confluence of unitary representations of states with the non-unitary 8-spinor reps in Sec. 2. The process yields a Dirac equation for 8-spinors that is shown to include a traditional Dirac equation for the donor 4-spinor. The recipient 4-spinor has an extra coordinate dependence not given to the donor 4-spinor. In Sec. 4, this dependence on position makes the 8-spinor current behave as the electromagnetic 4-vector potential with the 4-spinor donor field as source.

\section{Spacetime symmetries}

A Poincar\'{e} transformation $(\Lambda,b)$ applied to the coordinates $x^{\mu}$ = $\left(x,y,z,t\right)$ of Minkowski spacetime preserves the scalar products of displacements, $\delta x_{\sigma} \delta y^{\sigma}$ = $\eta_{\mu \sigma} \delta x^{\mu} \delta y^{\sigma},$ where $\delta x^{\mu}$ = $x^{\mu}_{2} - x^{\mu}_{1},$ with $\mu,\nu,... \in$ $\{1,2,3,4\}$ = $\{x,y,z,t\}.$ The spacetime metric $\eta_{\mu \nu}$ is diagonal with
\begin{equation} \label{eta}
 \eta_{11}  = \eta_{22}  =\eta_{33}  = +1 \quad ; \quad \eta_{44}  = -1  \quad .
\end{equation}
Vector indices are raised and lowered with $\eta,$ as in $\delta x_{\sigma}$ = $\eta_{\mu \sigma} \delta x^{\mu}.$ 

Space and time inversions also preserve the scalar product $\delta x_{\sigma} \delta y^{\sigma},$ but these will be considered separately as needed. 

The  Poincar\'{e} transformation $(\Lambda,b)$ is a Lorentz rotation-boost combination $\Lambda$ followed by a translation along a displacement $b^{\mu}.$ The transformation is assumed to be connected to the identity and can result from successive infinitesimal transformations in many ways. A linear representation $D(\Lambda,b)$ of a transformation in the Poincar\'{e} group can be put in the form,\cite{Tung Theorem 10.4 p182}
\begin{equation} \label{DLambdaB}
 D(\Lambda,b)  =  \exp{(-i b_{\mu} P^{\mu}})\exp(i\omega_{\mu \nu}  J^{\mu \nu}/2) \quad ,
\end{equation}
where the momenta $P^{\mu}$ generate translations and the angular momenta $J^{\mu \nu}$ generate rotations/boosts.  

Finite dimensional matrices $J^{\mu \nu}$ exist that generate representations of the Lorentz group of rotation-boosts. These generators obey the algebra\cite{Weinberg QTF I (2.4.12) p. 60, Tung (10.2-18) p 185}
\begin{equation} \label{PCrules0}
i[J^{\mu \nu},J^{\rho \sigma}] =   \eta^{\nu \rho} J^{\mu \sigma} -\eta^{\mu \rho} J^{\nu \sigma} - \eta^{\sigma \mu} J^{\rho \nu} + \eta^{\sigma \nu} J^{\rho \mu} \quad .
\end{equation}
The irreducible non-unitary representations are labeled by spin, two half-integral positive numbers $(A,B),$ with $A,B \in $ $\{0,1/2,1$,$3/2,...  \}.$\cite{Weinberg QTF I Sec. 5.6}

Finite dimensional matrices that represent the Poincar\'{e} group extend Lorentz reps to also include momentum matrices $P^{\mu}$ that generate the translations. The generators $P^{\mu}$ obey the algebra\cite{Weinberg QTF I (2.4.12) p. 60, Tung (10.2-18) p 185}
\begin{equation} \label{PCrules1}
i[P^{\mu},J^{\rho \sigma}] =   \eta^{\mu \rho} P^{\sigma} -\eta^{\mu \sigma} P^{\rho}  \quad ,
\end{equation}
\begin{equation} \label{PCrules2}
[P^{\mu},P^{\nu}] =   0  \quad .
\end{equation}
These commutation relations require that Poincar\'{e} matrix reps combine irreducible Lorentz reps with `linked' spins. 

Lorentz reps with spins $(A,B)$ and $(C,D)$ can be 'linked' when
 \begin{equation} \label{linked}
 C = A \pm 1/2 \quad ; \quad D = B \pm 1/2 \, ,
 \end{equation}
$A$ and $B$ differ from $C$ and $D$ by a half.
For spin 1/2, right and left 2-spinors, spin $(0,1/2)$ and $(1/2,0),$ 
can be linked one to the other. A left 2-spinor links to a right 2-spinor and vice versa. However, the linking must not be reflective; if a right and a left 2-spinor are each linked to the other, then the $P^{\mu}$ no longer commute, so (\ref{PCrules2}) is violated.\cite{Liubarsky, Shurtleff}

An 8-spinor $\psi$ consists equally of two right-handed 2-spinors and two left-handed ones. These can be organized into two distinct 4-spinors each transforming independent of the other under the Lorentz group of rotations and boosts. The rotation/boost generators are\cite{Jdef} 
\begin{equation} \label{8J}
J^{\mu \nu} = -\frac{i}{4}(\gamma^{\mu}\gamma^{\nu}-\gamma^{\nu}\gamma^{\mu}) \, , \quad \quad {\mathrm{where}} \quad  \gamma^{\mu} = \pmatrix{\gamma^{\mu}_{(4)} && 0 \cr 0 && \gamma^{\mu}_{(4)}}  \quad . 
\end{equation}
and $\gamma^{\mu}_{(4)}$ are a set of Dirac gamma matrices,
\begin{equation} \label{4gammas}
\gamma^{\mu}_{(4)}\gamma^{\nu}_{(4)}+\gamma^{\nu}_{(4)}\gamma^{\mu}_{(4)} = 2\eta^{\mu \nu} {\mathbf{1}}
\end{equation}
with ${\mathbf{1}}$ the unit matrix in 4-dimensions.

Translations link right and left 2-spinors. As noted previously, any right or left 2-spinor can be linked to any left or right 2-spinor, respectively. From among all the linking options with an 8-spinor, one can show that only linking one 4-spinor's right and left 2-spinors to the other 4-spinor's left and right 2-spinors preserves parity. We link lower to upper. Writing the $8 \times 8$ matrices in $2 \times 2$ and $4 \times 4$ blocks, the momentum  generators $P^{\mu}$ are 
\begin{equation} \label{Pmu}
P^{\mu} = i k\pmatrix{0 && 0 && 0 && 0 \cr 0 && 0 && 0 && 0 \cr 0 && -\sigma^{\mu} && 0 && 0 \cr  \sigma_{\mu} && 0 && 0 && 0 } =  \pmatrix{0 && 0  \cr P^{\mu}_{21} && 0 } \quad ,
\end{equation}
where all the nonzero terms are in the $4 \times4$ block $P^{\mu}_{21}.$ Applied to an 8-spinor $\psi$, the block $P^{\mu}_{21}$ links the lower 4-spinor $\psi_{5}$ to $\psi_{8}$ to the upper 4-spinor $\psi_{1}$ to $\psi_{4}.$

The momentum matrices (\ref{Pmu}) are written in a Weyl-representation with $4 \times 4$ gamma matrices\cite{WsGammas}
\begin{equation} \label{rep}
\gamma^{\mu}_{(4)} = i \pmatrix{0 && -\sigma^{\mu} \cr \sigma_{\mu} && 0}  \quad ,
\end{equation}
where the Pauli matrices are
\begin{equation} \label{sigma}
\sigma^{\mu} = \pmatrix{\delta^{\mu t} + \delta^{\mu z} && \delta^{\mu x} - i \delta^{\mu y} \cr \delta^{\mu x} + i \delta^{\mu y} && \delta^{\mu t} - \delta^{\mu z}} \quad {\mathrm{and}} \quad \sigma_{\mu} = \pmatrix{-\delta^{\mu t} + \delta^{\mu z} && \delta^{\mu x} - i \delta^{\mu y} \cr \delta^{\mu x} + i \delta^{\mu y} && -\delta^{\mu t} - \delta^{\mu z}} \quad , 
\end{equation} 
where $\delta^{\mu \nu}$ is one when $\mu$ = $\nu$ and zero when $\mu \neq$ $\nu.$ Note that raising and lowering indices with $\eta,$ (\ref{eta}), is equivalent to changing the signs of the time components. A Weyl-representation is convenient for separating left and right 2-spinors and discussing parity. 

One can show that the angular momentum matrices $J^{\mu \nu}$ in (\ref{8J}) and the momentum matrices (\ref{Pmu}) satisfy the commutation relations of the Poincar\'{e} algebra, eqns. (\ref{PCrules0},\ref{PCrules1},\ref{PCrules2}).

Spatial inversion, parity, exchanges right and left 2-spinors while leaving spin up and spin down invariant. In the Weyl-representation (\ref{rep}), spatial inversions are carried out by the the matrix $\beta,$ with\cite{WsBETA}
\begin{equation} \label{beta}
\beta = i\gamma^{4} = \pmatrix{0 && 1 && 0 && 0 \cr 1 && 0 && 0 && 0 \cr 0 && 0 && 0 && 1 \cr 0 && 0 && 1 && 0 }\quad ,
\end{equation}
where each entry is a $2 \times 2$ block. Clearly, $\beta$ exchanges right and left 2-spinors in both the upper 4-spinor and in the lower 4-spinor, but leaves spin up/down alone. In (\ref{beta}), a free multiplicative factor is adjusted so that $\beta$ is its own inverse, $\beta^{-1}$ = $\beta,$ or, put another way, $\beta^2$ = 1. It is straightforward to show that
\begin{equation} \label{JPparity}
\beta J^{\mu \nu} \beta = J_{\mu \nu} \quad {\mathrm{and}} \quad \beta P^{\mu} \beta = - P_{\mu}  \quad ,
\end{equation}
which are the appropriate changes in parity when the spacetime metric is the diagonal matrix $\eta$ = diag$(+1,+1,+1,-1).$

Because the momentum matrices $P^{\mu}$ are off-diagonal, one has $P^{\mu} P^{\nu}$ = 0, and the translation $D(1,b)$ simplifies,
\begin{equation} \label{DTrans}
D(1,b) = \exp{(-i b_{\mu} P^{\mu})} = 1 -i b_{\mu} P^{\mu} \quad .
\end{equation}
Thus translation matrices are linear in the displacements $b.$

With the matrices $P^{\mu}$ in the representation (\ref{Pmu}), translating an 8-spinor gives
$$
D(1,b) \psi = \left( 1 -i b_{\mu} P^{\mu} \right)\psi = $$
$$= \pmatrix{1 && 0 && 0 && 0 \cr 0 && 1 && 0 && 0 \cr 0 && -i k b_{\mu}\sigma^{\mu} && 1 && 0 \cr i k b_{\mu}\sigma_{\mu} && 0 && 0 && 1 } \pmatrix{\pmatrix{\psi_{1} \cr \psi_{2}} \cr \pmatrix{\psi_{3} \cr \psi_{4}} \cr \pmatrix{\psi_{5} \cr \psi_{6}} \cr \pmatrix{\psi_{7} \cr \psi_{8}}}
  = \pmatrix{\pmatrix{\psi_{1} \cr \psi_{2}} \cr \pmatrix{\psi_{3} \cr \psi_{4}} \cr \pmatrix{\psi_{5} \cr \psi_{6}} -i k b_{\mu}\sigma^{\mu} \pmatrix{\psi_{3} \cr \psi_{4}}  \cr \pmatrix{\psi_{7} \cr \psi_{8}}+i k  b_{\mu}\sigma_{\mu}\pmatrix{\psi_{1} \cr \psi_{2}}} \quad .
$$ 
The extra $()$ are for clarity. Thus components $\psi_{5}$ to $\psi_{8}$ receive a contribution from components $\psi_{1}$ to $\psi_{4}.$ In view of this, let us call the upper 4-spinor the donor 4-spinor $\psi_{\mathrm{donor}}$ and call the lower 4-spinor the receiver 4-spinor $\psi_{\mathrm{receiver}}$  
$$
\psi_{\mathrm{donor}} \equiv \pmatrix{\psi_{1} \cr \psi_{2} \cr \psi_{3} \cr \psi_{4}} \quad ; \quad  \psi_{\mathrm{receiver}} \equiv \pmatrix{\psi_{5} \cr \psi_{6} \cr \psi_{7} \cr \psi_{8}} \quad .
$$
The amount added to $\psi_{\mathrm{receiver}}$ is independent of $\psi_{\mathrm{receiver}}.$ This is like translating coordinates because no matter what the values of the coordinates of a point, a translation adds the same displacement.

\section{Spin half fields} \label{Shf}

Eight component annihilation fields $\psi^{+}_{l}(x)$ and creation fields $\psi^{-}_{l}(x)$ are sums of annihilation operators $a_{\sigma}({\mathbf{p}})$ and creation operators $a^{\dagger}_{\sigma}({\mathbf{p}})$ with coefficient functions $u_{l \sigma}(x;{\mathbf{p}})$ and $v_{l \sigma}(x;{\mathbf{p}})$ as follows:\cite{psiPLUSminus}
\begin{equation} \label{psi0}
\psi^{+}_{l}(x) = \sum_{\sigma} \int d^3 p \enspace u_{l \sigma}(x;{\mathbf{p}}) a_{\sigma}({\mathbf{p}})  \quad ,
\end{equation}
and
\begin{equation} \label{psi021}
\psi^{-}_{l}(x) = \sum_{\sigma} \int d^3 p \enspace v_{l \sigma}(x;{\mathbf{p}}) a^{\dagger}_{\sigma}({\mathbf{p}})  \quad .
\end{equation}
The symbol $\mathbf{p}$ denotes the space components of the momentum, $\{p^{x},$$p^{y},$$p^{z}\}.$ The spacial components $\mathbf{p}$ determine the time component $p^{\,t}$ = $(m^2+\mathbf{p}^2),$ which is required to be positive for the class of Poincar\'{e} transformations considered here, $p^{\,t} > 0.$ The particle field is the sum of the two fields,\cite{kappaLAMBDA} $\psi$ = $\psi^{+} + \psi^{-} \, .$ 

Suppose the unitary operator $U(\Lambda,b)$ acts on particle states when the system is boosted and rotated with the Lorentz transformation $\Lambda$ followed by a translation along the displacement $b.$ The operators transform as single particle states under $U$ and the coefficient functions $u$ and $v$ are invariant. 

Thus the creation and annihilation operators transform with unitary reps,\cite{Weinberg QTF I p. 193 (5.1.11) (5.1.12)}.
\begin{equation} \label{Da}
U(\Lambda,b) a({\mathbf{p}},\sigma) {U}^{-1}(\Lambda,b) = e^{i \Lambda p \cdot b} \sqrt{\frac{(\Lambda p)^t}{p^t}}  \sum_{\bar{\sigma}} D^{(j)}_{\sigma \bar{\sigma}}\left(W^{-1}\left(\Lambda,p\right) \right) a_{\bar{\sigma}}({\mathbf{ p_{\Lambda}}})   \quad 
\end{equation}
$$
U(\Lambda,b) a^{\dagger}({\mathbf{p}},\sigma) {U}^{-1}(\Lambda,b) = e^{-i \Lambda p \cdot b} \sqrt{\frac{(\Lambda p)^t}{p^t}}  \sum_{\bar{\sigma}} D^{(j) \, \ast}_{\sigma \bar{\sigma}}\left(W^{-1}\left(\Lambda,p\right) \right) a^{\dagger}_{\bar{\sigma}}({\mathbf{ p_{\Lambda}}})   \quad ,
$$
where $j$ is the spin of the particles and the space components of the transformed momentum $\Lambda p$ are denoted ${\mathbf{ p_{\Lambda}}}.$

The matrices $D^{(j)}$ form a  spin $j$ unitary representation of the Lorentz transformation $\Lambda$, while the phase factors $e^{\pm i p \cdot b}$ are the unitary representations of the translation by $b.$ Since the representation is unitary and boosts form a non-compact group, the rep is infinitely dimensional which shows up here in the dependence of the matrices $D^{(j)}$ on the particle's momentum $p.$ `$W$' stands for the Wigner rotation associated with the transformation $\Lambda$ for particle momentum $p.$ For details see reference \cite{Weinberg QTF I Chap5}.

One requires the fields $\psi^{+}_{l}(x)$ and $\psi^{-}_{l}(x)$ to transform differently. Fields transform with a nonunitary representation $D(\Lambda,b)$ as in (\ref{DLambdaB}),
 \begin{equation} \label{Upsi+U=Dm1psi}
U(\Lambda,b) \psi^{\pm}_{l}(x) {U}^{-1}(\Lambda,b) =  D^{-1}_{l \bar{l}}(\Lambda,b)  \psi^{\pm}_{\bar{l}}(\Lambda x + b) \quad .
\end{equation}
These nonunitary transformations differ from those in Ref. \cite{Weinberg QTF I Chap5} because we include the same translation along a displacement $b$ that is applied in the unitary transformations of the operators $a$ and $a^{\dagger}.$ This is the reason for writing this paper.

The contrary transformations of the fields and operators constrains the  coefficient functions. By (\ref{psi0}, \ref{psi021}, \ref{Da}, \ref{Upsi+U=Dm1psi}), one can show that 
\begin{equation} \label{Du2h}
  u_{l \sigma}(x,{\mathbf{p}}) = \sqrt{\frac{ m}{p^t}}\, e^{i p \cdot x} D_{l \bar{l}}(L(p),x) u_{\bar{l} \sigma}(0,{\mathbf{0}}) \quad \end{equation}
$$
  v_{l \sigma}(x,{\mathbf{p}}) = \sqrt{\frac{ m}{p^t}} \, e^{-i p \cdot x} D_{l \bar{l}}(L(p),x) v_{\bar{l} \sigma}(0,{\mathbf{0}}) \quad,
$$ 
where the quantities $u_{\bar{l} \sigma}(0,{\mathbf{0}})$ and $v_{\bar{l} \sigma}(0,{\mathbf{0}})$ are evaluated at the origin $x$ = 0 with the particle at rest, 3-momentum ${\mathbf{ p}}$ =  ${\mathbf{ 0}}$ and 4-momentum $k$ = $(0,0,0,m).$ 

 The `standard boost', the Lorentz transformation $L(p),$ is given by\cite{W QTF I p 68 L(p)}
\begin{equation} \label{L(p)}
L^{i}_{k}(p) = \delta^{i}_{k} + (1+\gamma)^{-1} m^{-2} p^{i}p^{k}\, ,
\end{equation}
 \begin{equation} \label{L} 
 L^{i}_{4} = L^{4}_{i} = m^{-1}p^{i} \quad {\mathrm{and}} \quad L^{4}_{4} = \gamma = m^{-1} p^{4}\, ,
\end{equation} 
 transforms the rest frame of the particle with 4-momentum   $k^{\mu}$ = $(0,0,0,M)$  to a frame with $p^{\mu}$ = $({\mathbf{p}},\sqrt{1-{\mathbf{p}}^2}),$ a rotation taking $\hat{p}$ to $\hat{z}$ followed by a boost along $z$ followed by a rotation taking the unit vector $\hat{z}$ to $\hat{p}.$ The derivation of (\ref{Du2h}) follows the procedures in reference \cite{Weinberg QTF I Chap5}.

 Note that the coefficient functions are not simple plane waves because of the coordinate dependence in the translation part of the transformation matrix $D_{l \bar{l}}(L(p),x).$ By rewriting (\ref{Du2h}) as
\begin{equation} \label{Du2ha}
  u_{m \sigma}(x,{\mathbf{p}}) = e^{i p \cdot x} D_{m n }(1,x)u_{n \sigma}(0,{\mathbf{p}}) \quad ,
\end{equation}
\begin{equation} \label{Du2h1a}
  v_{m \sigma}(x,{\mathbf{p}}) = e^{-i p \cdot x} D_{m n }(1,x)v_{n \sigma}(0,{\mathbf{p}}) \quad ,
\end{equation}
we can collect together the coordinate independent parts, $u_{n \sigma}(0,{\mathbf{p}})$ and $v_{n \sigma}(0,{\mathbf{p}}),$ of the coefficient functions. 
 
 In (\ref{Da}), the matrices $D^{(j)}$ form a unitary representation of the Lorentz group that is applied to particle states and annihilation operators. These generate rotations not boosts by the Wigner rotation mechanism. One can show by Schur's lemma\cite{H1} that $j$ = 1/2 and we can take   
\begin{equation} \label{Du2e}
    J^{(j)k} = \frac{1}{2}\sigma^{k}   \quad ,
\end{equation}
without loss of generality. Another consequence is that the four 2-spinors that make up the 8-spinor coefficient functions when they are evaluated at the origin in the particle rest frame each commute with the three Pauli matrices $\sigma^{k}.$ Each 2-spinor is therefore proportional to the unit $2 \times 2$ matrix. Thus, we write $u_{m \sigma}(0,{\mathbf{0}})$ = $c_{a} \delta_{[m \; {\mathrm{mod2}}] \sigma}, $ where $a$ = 1 for the first 2-spinor with $m$ = $(1,2),$ $a$ = 2 for $m$ = $(3,4),$ $a$ = 3 for 8-spinor indices $m$ = $(5,6)$ and $a$ = 4 for indices $(7,8).$ One has
\begin{equation} \label{Du2f}
  u_{l, \, +1/2}(0,{\mathbf{0}}) =   \pmatrix{c_1 \cr 0 \cr c_2 \cr 0 \cr c_3 \cr 0 \cr c_4 \cr 0}    \quad \quad   u_{l, \, -1/2}(0,{\mathbf{0}}) =   \pmatrix{0 \cr c_1 \cr 0 \cr c_2 \cr 0 \cr c_3 \cr 0 \cr c_4 } \quad ,
\end{equation}
where $l$ runs from 1 to 8 and the two values of the index $\sigma$ are $\pm 1/2.$ The four constants $c_{i}$ are arbitrary. For details see reference \cite{Weinberg QTF I p220 221} where the same procedure is applied to 4-spinors.

The constants $c_{i}$ may be constrained by their behavior under spatial inversion. As discussed in some detail in Ref. \cite{Wchapt5.5}, if the description is to be parity invariant, then the coefficient functions $u_{l, \, +1/2}(0,{\mathbf{0}})$ and $u_{l, \, -1/2}(0,{\mathbf{0}})$ should be eigenspinors of the parity-changing matrix $\beta,$
\begin{equation} \label{eigenBETA0}
\beta u_{n, \, \sigma}(0,{\mathbf{0}}) = u_{n, \, \sigma}(0,{\mathbf{0}}) \quad {\mathrm{and}} \quad \beta v_{{n} \sigma}(0,{\mathbf{0}}) = -v_{{n} \sigma}(0,{\mathbf{0}}) \quad ,
\end{equation}
where $\sigma$ = $\pm 1/2.$ For the Weyl-rep here with the $\beta$ in (\ref{beta}), the coefficient functions in (\ref{Du2f}) have constants $c_{i}$ constrained by $c_{1}$ = $c_{2}$ and $c_{3}$ = $c_{4}.$ Similar remarks apply to the coefficient functions $v_{n, \, \sigma}(0,{\mathbf{0}}).$ 

For a particle perhaps not at the origin and not at rest, one can apply the transformation $D(L(p),x).$  By (\ref{Du2h}) and (\ref{eigenBETA0}) and replacing $\beta$ with $i\gamma^{4}$, one finds that the coefficient function $u_{l \sigma}(x,{\mathbf{p}})$ satisfies 
\begin{equation} \label{1221o}
  \left[ iD(L(p),x)\gamma^{4} D^{-1}(L(p),x)\right]_{ln} u_{n \sigma}(x,{\mathbf{p}}) = u_{l \sigma}(x,{\mathbf{p}}) \quad .
\end{equation} 
By (\ref{DLambdaB}), $\gamma^{4}$ is transformed by $D(L(p),x)$ = $D(1,x)D(L(p),0),$ a Lorentz rotation/boost followed by a translation. 

Take $L(p)$ first, i.e. $D(L(p),0)\gamma^{4} D^{-1}(L(p),0).$ One can show by (\ref{PCrules0}) that vector matrices such as $\gamma^{\mu}$ transform with a Lorentz rotation/boost both as a vector and a second-order tensor. For $D(L(p),0)$ this means that 
\begin{equation} \label{vecTEN}
D(L(p),0)\gamma^{4} D^{-1}(L(p),0) = (L^{-1}(p))^{ 4}_{\nu} \gamma^{\nu} = -\frac{ 1}{m} \gamma^{\nu}p_{\nu} \quad ,
\end{equation}
where the component $(L^{-1}(p))^{ 4}_{\nu}$ = $ -p_{\nu} /m,$ by (\ref{L}) with $p \rightarrow$ $(-p^{i},p^{4})$ to get the inverse transformation. Now applying the translation transformation $D(1,x)$ = $\left( 1 -i x_{\mu} P^{\mu} \right)$ and its inverse to (\ref{1221o}) yields 
\begin{equation} \label{DgammaDuu}
-i\left[ \gamma^{\nu} - ix_{\mu}\left(P^{\mu}\gamma^{\nu} - \gamma^{\nu}P^{\mu}  \right)\right]_{kl} p_{\nu}  u_{l \sigma}(x,{\mathbf{p}}) = m u_{k \sigma}(x,{\mathbf{p}}) \quad 
\end{equation}
$$
-i\left[ \gamma^{\nu} - ix_{\mu}\left(P^{\mu}\gamma^{\nu} - \gamma^{\nu}P^{\mu}  \right)\right]_{kl} p_{\nu}  v_{l \sigma}(x,{\mathbf{p}}) = -m v_{k \sigma}(x,{\mathbf{p}}) \quad ,
$$
which is the Dirac equation at the coefficient function level for the 8-spinor field. 

By the $\gamma^{\mu}$ in (\ref{8J}) and the $P^{\mu}$ matrices in (\ref{Pmu}), the matrix $P^{\mu}\gamma^{\nu} - \gamma^{\nu}P^{\mu}$ has just one nonzero off-diagonal $4 \times 4$ block,
\begin{equation} \label{commPgamma}
P^{\mu}\gamma^{\nu} - \gamma^{\nu}P^{\mu}  = \pmatrix{0 && 0 \cr P^{\mu}_{21} \gamma^{\nu}_{(4)} -  \gamma^{\nu}_{(4)} P^{\mu}_{21} && 0 } \quad .
\end{equation}
The two zero blocks in the first row show that the matrix has no effect on the first 4 components of $u$ and $v,$ $ \{u_{1 \sigma},u_{2 \sigma},u_{3 \sigma},u_{4 \sigma}\}$ = $u^{\mathrm{donor}}_{i \sigma}$ and $ \{v_{1 \sigma},v_{2 \sigma},v_{3 \sigma},v_{4 \sigma}\}$ = $v^{\mathrm{donor}}_{i \sigma}.$ 

It follows that, when restricted to the donor 4-spinor coefficient function $u^{\mathrm{donor}}_{i \sigma},$ the Dirac equation (\ref{DgammaDuu}) simplifies to
\begin{equation} \label{DgammaDuuUP}
-i\gamma^{\nu}_{(4)\, ij} p_{\nu} \, u^{\mathrm{donor}}_{j \sigma}(x,{\mathbf{p}}) = mu^{\mathrm{donor}}_{i \sigma}(x,{\mathbf{p}}) \quad 
\end{equation}
$$
-i\gamma^{\nu}_{(4) \, ij} p_{\nu} \, v^{\mathrm{donor}}_{j \sigma}(x,{\mathbf{p}}) = -mv^{\mathrm{donor}}_{i \sigma}(x,{\mathbf{p}}) \quad .
$$
By (\ref{Du2h}), the coefficient function $u^{\mathrm{donor}}_{i \sigma}(x,{\mathbf{p}})$ depends on $x$ only in the phase factor $\exp{(i p \cdot x)}.$ Thus, the momentum factor $p_{\nu}$ can be obtained with the differential $-i \partial_{\mu}$ and multiplying $u^{\mathrm{donor}}_{i \sigma}(x,{\mathbf{p}})$ by $p_{\nu}$ is equivalent to applying $-i \partial_{\mu}.$ For $v^{\mathrm{donor}}$ its $+i \partial_{\mu},$ since  $v^{\mathrm{donor}}$ has $\exp{(-i p \cdot x)}$ in (\ref{Du2h}). One finds that
\begin{equation} \label{pdmu}
   - \gamma^{\nu} _{(4)}\partial_{\nu}  u^{\mathrm{donor}}_{ \sigma}(x,{\mathbf{p}})  = mu^{\mathrm{donor}}_{ \sigma}(x,{\mathbf{p}})  \quad 
\end{equation}
$$
   - \gamma^{\nu}_{(4)} \partial_{\nu}  v^{\mathrm{donor}}_{ \sigma}(x,{\mathbf{p}})  = mv^{\mathrm{donor}}_{ \sigma}(x,{\mathbf{p}})  \quad ,
$$
where some indices are dropped to avoid clutter. The differential momentum operator $\partial_{\nu}$ is linear and can be applied to all terms of the expansion of the donor 4-spinors $\psi^{+}_{l}(x)$ and $\psi^{-}_{l}(x)$ in (\ref{psi0}) and (\ref{psi021}), so we get
\begin{equation} \label{Dirac}
- \gamma^{\nu}_{(4)}  \partial_{\nu}  \, \psi^{ {\mathrm{donor}}}(x) = m \psi^{{\mathrm{donor}}}(x) \quad ,
\end{equation}
which includes the $\psi^{+}(x)$ and $\psi^{-}(x)$ fields. Thus the donor 4-spinor field $\psi^{\mathrm{donor}}$ = $(\psi_{1} , \psi_{2} ,$ $ \psi_{3} , \psi_{4})$ obeys the standard Dirac equation for a free spin 1/2 particle with mass $m.$ 
   
The situation for the receiver 4-spinor field $\psi^{\mathrm{receiver}},$ 8-spinor components 5 through 8, is complicated because the coefficient functions $u_{l \sigma}(x,{\mathbf{p}})$ are not simple plane waves like $\exp{(\pm i x_{\mu}p^{\mu})}$, but also depend on the coordinates $x$ via the translation matrix $D(1,x)$ in (\ref{Du2h}). 

Having found the usual Dirac equation for the donor 4-spinor, we choose to avoid discussing the Dirac equation of the receiver 4-spinor and concentrate on another aspect of its coordinate dependence.

\section{The E-M vector potential} \label{SecLEC}

The position dependence of the coefficient functions $u_{l \sigma}(x,{\mathbf{p}})$ and $v_{l \sigma}(x,{\mathbf{p}})$  in (\ref{Du2h}) is complicated by the presence of both phase factors $\exp{(\pm i x_{\mu}p^{\mu})}$ and the translation matrix $D(1,x)$. At the finish of the preceding section, by considering just the donors, $u^{\mathrm{donor}}_{l \sigma}(x,{\mathbf{p}})$ and $v^{\mathrm{donor}}_{l \sigma}(x,{\mathbf{p}}),$ we avoided the translations $D(1,x).$ In this section, we avoid the phase factors $\exp{(\pm i x_{\mu}p^{\mu})}$ by calculating currents.

Define the current $j^{\mu}$ of a coefficient function $u_{n \sigma}(x,{\mathbf{p}})$ to be
\begin{equation} \label{CVP1}
  j^{\mu}(x;{\mathbf{p}},\sigma) = N_{8} \enspace \bar{u}_{n \sigma}(x,{\mathbf{p}})\gamma^{\mu}_{nm} u_{m \sigma}(x,{\mathbf{p}}) \quad ,
\end{equation}
where $N_{8}$ is a normalization constant for the 8-spinor, there is no sum over $\sigma$ and $\bar{u}$ = $u^{\dagger} \gamma^{4}.$ Thus there is one current for each momentum ${\mathbf{p}}$ and each spin $\sigma$ = $\pm 1/2.$ The currents for the coefficients $v$ are similar. 

Separating the coordinate and coordinate-independent parts, as in (\ref{Du2ha}), we have 
\begin{equation} \label{CVP1a}
  j^{\mu}(x;{\mathbf{p}},\sigma) = N_{8} \enspace \bar{u}_{n \sigma}(0,{\mathbf{p}}) \left[ \gamma^{4} D^{\dagger}(1,x)   \gamma^{4} \gamma^{\mu}  D(1,x) \right]_{nm} u_{m \sigma}(0,{\mathbf{p}}) \quad ,
\end{equation}
with all the $x$-dependence confined to the translation $D(1,x)$ and its conjugate in the square brackets.

By (\ref{Pmu}) and (\ref{DTrans}), the translation $D(1,x)$  = $\exp{(-i x_{\beta} P^{\beta})}$ = $1 -i x_{\beta} P^{\beta}$ is linear in $x.$ It follows that $j^{\mu}(x;{\mathbf{p}},\sigma)$ is quadratic in $x.$  One finds that the  part of the current that is of second order in $x$ is given for the $P^{\mu}$ in (\ref{Pmu}) in the Weyl-rep (\ref{rep}) by 
 \begin{equation} \label{quadPART}
   j^{\mu}(x;{\mathbf{p}},\sigma)  \stackrel{\;\;x^2}{\asymp} N_{8} \enspace \bar{u}_{n \sigma}(0,{\mathbf{p}}) \left[ k^{2} x_{\alpha}x_{\beta}  \pmatrix{\gamma^{\alpha}_{(4)} \gamma^{\mu}_{(4)} \gamma^{\beta}_{(4)} && 0 \cr 0 && 0 }\right]_{nm} u_{m \sigma}(0,{\mathbf{p}}) \quad , 
\end{equation}
where ``$\stackrel{\;\;x^2}{\asymp}$'' means that only terms of order $x^{2}$ are displayed. The $4\times4$ donor-donor block of the square brackets has all the nonzero matrix components. 

By (\ref{quadPART}), only the donor-donor terms, $\bar{u}^{\mathrm{donor}}\left[k^{2} x_{\alpha}x_{\beta}  \gamma^{\alpha}_{(4)} \gamma^{\mu}_{(4)} \gamma^{\beta}_{(4)} \right]u^{\mathrm{donor}},$ have nonzero second order derivatives. Thus 
\begin{equation} \label{d2j}
 \partial_{\alpha} \partial_{ \beta}j^{\mu}(x;{\mathbf{p}},\sigma) =  N_{8} k^{2} \bar{u}^{\mathrm{donor}}_{i \sigma}(0,{\mathbf{p}})\left[  \gamma_{\alpha}^{(4)} \gamma^{\mu}_{(4)} \gamma_{\beta}^{(4)} + \gamma_{\beta}^{(4)} \gamma^{\mu}_{(4)} \gamma_{\alpha}^{(4)}\right]_{ij} u^{\mathrm{donor}}_{j \sigma}(0,{\mathbf{p}}) \quad ,
\end{equation}
where $\gamma_{\alpha}^{(4)} $ = $\eta_{\alpha \rho}\gamma^{\rho}_{(4)}$ and $i,j \in$ $\{1,2,3,4\}$ denoting the components of the 4-spinor donor portion $u^{\mathrm{donor}}_{i \sigma}$ of the 8-spinor $u_{n \sigma}.$ 

From (\ref{d2j}) and the identity $\gamma^{\alpha}_{(4)} \gamma^{\mu}_{(4)} \gamma_{\alpha}^{(4)}$ = $2 \gamma^{\mu}_{(4)} ,$ we get
\begin{equation} \label{boxj1} 
 \eta^{\alpha \beta}\partial_{\alpha} \partial_{ \beta}j^{\mu}(x;{\mathbf{p}},\sigma) = 2 N_{8} k^{2} \bar{u}^{\mathrm{donor}}_{i \sigma}(0,{\mathbf{p}})\left[  \gamma^{\alpha}_{(4)} \gamma^{\mu}_{(4)} \gamma_{\alpha}^{(4)} \right]_{ij} u^{\mathrm{donor}}_{j \sigma}(0,{\mathbf{p}}) \quad 
\end{equation}
$$
 = 4 N_{8} k^{2} \bar{u}^{\mathrm{donor}}_{i \sigma}(0,{\mathbf{p}}) \gamma^{\mu}_{(4) \, ij} u^{\mathrm{donor}}_{j \sigma}(0,{\mathbf{p}})
$$
Defining the donor 4-vector current by
$$  j_{{\mathrm{donor}}}^{\mu}({\mathbf{p}},\sigma) \equiv  N_{4} \enspace \bar{u}^{\mathrm{donor}}_{i \sigma}(0,{\mathbf{p}})\gamma^{\mu}_{(4) \, ij} u^{\mathrm{donor}}_{j \sigma}(0,{\mathbf{p}}) \quad 
$$
yields, by (\ref{boxj1}),
\begin{equation} \label{boxj} 
\partial^{\alpha} \partial_{\alpha}j^{\mu}(x;{\mathbf{p}},\sigma) =  4{k}^2 \frac{N_{8}}{N_{4}} j_{{\mathrm{donor}}}^{\mu}({\mathbf{p}},\sigma) \quad .
\end{equation}
The quantity $N_{4}$ is the normalization constant for the 4-spinor $u^{\mathrm{donor}}.$ 

Similar manipulations show that 
\begin{equation} \label{2ndTermInVectorPot} 
\partial^{\mu} \partial_{\beta}j^{\beta}(x;{\mathbf{p}},\sigma) =  -8{k}^2 \frac{N_{8}}{N_{4}} j_{{\mathrm{donor}}}^{\mu}({\mathbf{p}},\sigma) \quad .
\end{equation}

The donor current $j_{{\mathrm{donor}}}^{\mu}$ is independent of position $x,$ because the four donor coefficient functions, $u^{\mathrm{donor}}$ = $(u_{1},u_{2},u_{3},u_{4}),$ do not receive any position-dependent terms from the translation matrix $D(1,x)$ in (\ref{Du2h}) and phase factors $\exp{(\pm i px)}$ cancel in the current. Thus, all the position $x$ dependence of the current $j^{\mu}(x;{\mathbf{p}},\sigma)$ is embedded with the four receiver coefficients, $u^{\mathrm{receiver}}$ =  $(u_{5},u_{6},u_{7},u_{8}).$

Subtracting (\ref{2ndTermInVectorPot}) from (\ref{boxj}) gives
\begin{equation} \label{jMaxwell}
  \partial^{\alpha} \partial_{\alpha} j^{\mu}(x;{\mathbf{p}},\sigma) -\partial^{\mu} \partial_{\beta} j^{\beta}(x;{\mathbf{p}},\sigma) = 12{k}^2 \frac{N_{8}}{N_{4}}   j_{{\mathrm{donor}}}^{\mu}({\mathbf{p}},\sigma) \quad .
\end{equation}
Comparing this to Maxwell's source equation shows that the 8-spinor current $j^{\mu}(x;{\mathbf{p}},\sigma)$ is proportional to an electromagnetic vector potential $a^{\mu},$
 \begin{equation} \label{CVP9b}
 a^{\mu} \equiv q \frac{N_{4}}{12{k}^2 N_{8}} j^{\mu}(x;{\mathbf{p}},\sigma) \quad .
 \end{equation}
The source of the electromagnetic vector potential $a^{\mu}(x;{\mathbf{p}},\sigma)$ is the 4-spinor donor current $j_{{\mathrm{donor}}}^{\mu}$ with a charge $q$ and mass $m$ and restricted to a state with momentum $\mathbf{p}$ and spin $\sigma.$ By (\ref{jMaxwell}) and (\ref{CVP9b}), we have
\begin{equation} \label{CVP9c}
  \partial^{\alpha} \partial_{\alpha} a^{\mu}(x;{\mathbf{p}},\sigma) -\partial^{\mu} \partial_{\beta} a^{\beta}(x;{\mathbf{p}},\sigma) = q  j_{{\mathrm{donor}}}^{\mu}({\mathbf{p}},\sigma) \quad ,
\end{equation}
which is coincident with Maxwell's source equation.\cite{Maxwell,Maxwell2} 

It may be important to include translations in quantum field theory because the group of translations is a fundamental spacetime symmetry group and its inclusion may provide insight into the force fields produced by spin 1/2 sources.

\end{document}